\documentclass[twocolumn,secnumarabic,amssymb, nobibnotes, aps, pre, superscriptaddress]{revtex4-1}

\usepackage{float}
\usepackage{graphicx}
\usepackage{subfigure}
\usepackage{amsmath} 
\usepackage{epstopdf}
\usepackage{setspace}
\usepackage[shortlabels]{enumitem}
\usepackage{mathrsfs}
\usepackage{subfig}
\usepackage{color,soul}
\usepackage[dvipsnames]{xcolor}
\usepackage{dcolumn,amssymb,subfig}
\usepackage{comment}
\usepackage{upgreek}

\renewcommand{\vec}[1]{\boldsymbol{#1}}
 \newcommand{\tens}[1]{\boldsymbol{#1}}

\newcommand*\diff{\mathop{}\!\mathrm{d}}

\begin{document}


\title{Activity-induced instabilities of brain organoids}


\author{Kristian Thijssen}
\thanks{K.T. contributed equally to this work with G.K.}
\affiliation{Yusuf Hamied Department of Chemistry, University of Cambridge, Lensfield Rd, Cambridge CB2 1EW, UK}
\author{Guido L. A. Kusters}
\thanks{G.L.A.K. contributed equally to this work with K.T.}
\affiliation{ Department of Applied Physcis, Eindhoven University of Technology, Eindhoven, The Netherlands}
\author{Amin Doostmohammadi}
\email{doostmohammadi@nbi.ku.dk}
\affiliation{ The Niels Bohr Institute, University of Copenhagen, Copenhagen, Denmark}


\begin{abstract}
We present an analytical and numerical investigation of the activity-induced hydrodynamic instabilities in model brain organoids. While several mechanisms have been introduced to explain the experimental observation of surface instabilities in brain organoids, the role of activity has been largely overlooked. Our results show that the active stress generated by the cells can be a, previously overlooked, contributor to the emergence of surface deformations in brain organoids.
\end{abstract}

\maketitle


\maketitle

\section{Introduction}
The surface of the human brain is characterised by a complex pattern of folds (\textit{gyri}) and troughs (\textit{sulci}), allowing for a high area-to-volume ratio \cite{ronan2013differential,tallinen2014gyrification,tallinen2016growth}. 
This intricate structure has increasingly been linked to intellectual ability, marking reduced cortical folding as indicative of cerebral impairments such as ``smooth brain" (\textit{lissencephaly}).
Indeed, afflicted human brains exhibit a markedly lower degree of gyrification, resulting in reduced life expectancy and intellectual disability \cite{reiner1993isolation,reiner2013lis1}.
Recent advancements in the field of stem-cell research provide a controlled, \textit{in vitro} model system allowing the study of gyrification in the form of brain organoids: cultured, three-dimensional arrangements of pluripotent stem cells replicating some of the key features of human brain development \cite{lancaster2013cerebral,pacsca2015functional}. 
Although the brain organoid model has been widely used in furthering understanding of a wealth of diseases \cite{clevers2016modeling,bershteyn2017human,di2017use}, the underlying physical mechanism governing gyrification has yet to be pinned down decisively. 

\begin{figure} 
    \centering
    \includegraphics[width=0.35\textwidth]{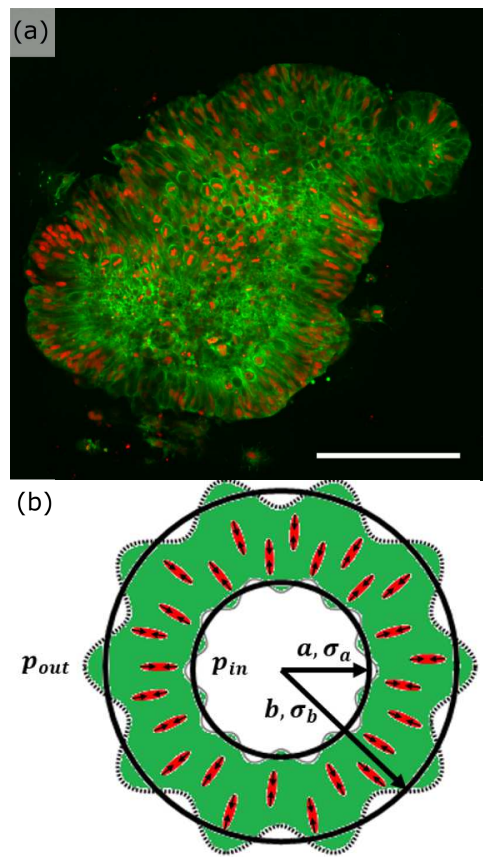}
    \caption{
\textbf{Depiction of brain organoids observed in experiments and the active nematic model we use to illustrate the role of contractile stress in folding. }
(a) Fluorescence image during organoid growth. Actin filaments found in the organoid cells are coloured green, and the cell nuclei are coloured red. Figure adapted from~\cite{karzbrun2018human} and the scale bar is $200\mu m$.
(b) Schematic representation of the theoretical model.  Filled red ellipsoids represent cell alignment, and the corresponding black arrows indicate contractile stress. The model assumes uniform active stress, which is sufficient to capture the folding of the interfaces. The remainder of the organoid is coloured green to represent the organoid cells. Solid black lines indicate undeformed organoid (low activity) and dashed black lines indicate a deformed organoid (high activity) due to the active  stress.
}
\label{fig:organoidmodel}
\end{figure}

In this context, recent experimental work by Karzbrun \textit{et al.} broke new ground by experimentally probing brain organoids for the onset of an interface instability that results in the formation of folds on the organoid \cite{karzbrun2018human} (see Figure \ref{fig:organoidmodel}a). 
In particular, they show the development of brain organoids over the course of several days, observing the self-organisation of a concentric shell of cells around a spherical cavity (\textit{lumen}). 
Due to their active motility, the cell nuclei continually move radially inward and outward, dividing at the inner surface and eventually accumulating at the outer surface. 
They find prior to the onset of mechanical instability an increase in the density of the nuclei, as well as the aspect ratio, which indicates compression. 
Taking inspiration from polymer gel models~\cite{tanaka1987mechanical}, they subsequently argue that these instabilities emerge from the \textit{differential swelling} of the inner and outer cortex, inducing compressive stress \cite{ronan2013differential,tallinen2016growth}.

Building on these observations, Balbi \textit{et al.}~\cite{balbi2020mechanics} showed that an interplay between the lumen compression and the remodelling of the cortex determines the interface instability of the organoids. Riccobelli and Bevilasqua further showed~\cite{riccobelli2020} that surface tension generated by intercellular adhesion in cellular aggregates also contributes to determining the onset of the interface instability. Recently, Engstrom et al.~\cite{engstrom2018} argued that such elastic instabilities based on differential growth sketch an incomplete picture of the  folding of brain organoids. Instead, they introduce a system-spanning fibrous model of organoids with an elastic core ensnared by a growing, fluid-like film, suggesting that the details of the microstructure play an important role in the emergence and structure of the wrinkles.

Notwithstanding these important contributions, here we consider a hitherto overlooked aspect of the microstructural complexity of brain organoids in active stress generation by the cells, showing that activity can induce folding at the surface of model organoids in the form of hydrodynamic instabilities. Active stress between cells has been shown to induce surface instabilities in epithelial cell layers \cite{hannezo2012mechanical,hannezo2014theory} and surface deformations in membranes \cite{simon2018direct}.
The model presented here expands on this list and is motivated by experiments in which the brain organoids are treated with cytoskeletal-inhibiting drugs~\cite{straight2003dissecting}, where a marked decrease in the number of folds exhibited and the sharpness of the folds are observed \cite{karzbrun2018human}. 
Since the cytoskeletal filaments inside the cells continuously generate active  stress, we conjecture that in addition to a purely elastic wrinkling phenomenon following differential swelling, the activity-driven instability can also contribute to the observed folding behaviour.

The concept of active stress generation by cells is well established in the context of active matter, which describes a material class that exists far from thermodynamic equilibrium by virtue of a local supply of energy. 
This energy is then converted into work by the constituents, resulting in active stress \cite{ramaswamy2010mechanics}.
Examples exude biological systems such as  actin filaments \cite{chakrabarti2007isotropic}, microtubule bundles \cite{sanchez2012spontaneous} and eukaryotic epithelial cells \cite{poujade2007collective,petitjean2010velocity}. Furthermore, recent discrete numerical modelling of epithelial shells~\cite{rozman2020collective} has demonstrated the role of activity in inducing morphological changes of the interface by considering the cell-cell tension in the apical  and basal surface planes of the cells.

Here, we present a generic, continuum, two-phase framework, in which we describe the brain organoid as an active gel, to study the spatiotemporal dynamics of active, self-deforming surfaces. This generic approach allows us to effectively describe organisation of cells on long time and length scales in terms of the hydrodynamics of the active gel, as well as include the orientational order and contractile activity of the cell cytoskeleton, without reference to the microscopic details \cite{zumdieck2008spontaneous}.
The hydrodynamic nature of the model, that accounts for the flow of cells, together with the explicit role of the contractile active stress generation by the cell cytoskeleton, is what sets it apart from those mentioned above.

We perform coarse-grained, continuum simulations in two-dimensions, showcasing that the number of folds is dependent on active stress generation by the contractile actomyosin machinery of the cells.  
An activity threshold needs to be surpassed for folding to occur, similar to experiments where the organoid needs to surpass a critical nuclear cell density for folding to occur. In this vein, our model naturally incorporates the effect of cytoskeletal-inhibiting drugs on the folding behaviour of the organoid, whereas alternative approaches, e.g., based on differential swelling, generally require the assumption of an actively contracted organoid core \cite{balbi2020mechanics}.
Additionally, we present a linear stability analysis of the governing equations in order to characterise the onset of the activity-driven instabilities. 
Taken together, our results suggest that active stress generation provides a currently-overlooked mechanism for cortical folding that is complementary to existing models.

\section{Model}

We model the brain organoid (see Figure \ref{fig:organoidmodel}) as a ring of an active gel (with inner radius $a$ and outer radius $b$)~\cite{prost2015active}, representing the cortex, surrounding a passive isotropic cavity (within the inner interface at radius $a$), which represents the lumen. 
Here, we differentiate between the cortex and lumen region by introducing a binary order parameter $\phi$, with $\phi\sim1$ in the cortex and $\phi\sim0$ within the lumen region. Within this cortex region, the cells and their associated cytoskeletal filaments are extended radially with clear orientational order (see Figure \ref{fig:organoidmodel}a). To account for the orientational order associated with this microstructural feature of the cortex, we introduce a nematic order parameter $\tens{Q}$ that is a symmetric, traceless tensor $Q_{\alpha\beta}=S\left(q_\alpha q_\beta-\frac{1}{2}\delta_{\alpha\beta}\right)$ \cite{doi2013soft}, where $S$ represents the magnitude of the orientational order, and $q$ indicates direction~\cite{prost1995physics}.
The use of this nematic order parameter is well established in the study of cellular and subcellular systems with elongated constituents such as the cell cytoskeleton~\cite{kumar2018tunable,maroudas2020topological} and confluent tissues of epithelial or fibroblast cells~\cite{duclos2018spontaneous,BlanchMercader2018}. By employing this nematic order parameter, our formulation allows for the dynamics of cell alignment within the cortex to be explicitly accounted for. Furthermore, this mapping allows us to describe the cortex as an active nematic system, for which the continuum equations are well established in the literature \cite{doostmohammadi2018active}. 

Within the active nematic framework we evolve the binary order parameter $\phi$ and the orientational order parameter $\tens{Q}$ through the Cahn-Hilliard~\cite{Cahn1958} and the Beris-Edwards equations~\cite{beris1994thermodynamics,giomi2012banding}, respectively:
\begin{eqnarray}
\partial_t \phi + \partial_k \left(v_k\phi\right) &= \mu,
\label{eqn:phi}\\
\left(\partial_t+v_k\partial_k\right)Q_{ij}-\Tilde{S}_{ij} &= H_{ij},
\label{eqn:Q}
\end{eqnarray}
where $\vec{v}$ is the velocity that advects order parameter fields, and $\tens{\Tilde{S}}$ represents the co-rotation terms describing the response of elongated particles to velocity gradients. The latter is given by
\begin{equation}\label{corotation}
\begin{split}
    \Tilde{S}_{ij}=\xi E_{ij}+Q_{ik}\Omega_{kj}-\Omega_{ik}Q_{kj},
\end{split}
\end{equation}
with the strain rate tensor $E_{ij}=\left(\partial_iv_j+\partial_jv_i\right)/2$ and the vorticity tensor $\Omega_{ij}=\left(\partial_iv_j-\partial_jv_i\right)/2$ describing the symmetric and asymmetric parts of the velocity gradient tensor, respectively. In addition, $\xi$ denotes a flow-alignment parameter, which determines the collective response of the orientation field to gradients in the velocity field.
These relaxation dynamics are governed by minimising the free energy of the system with respect to $\phi$ and $\tens{Q}$ through the normalised chemical potential 
\begin{align}
\mu &= \Gamma_\phi\left( \frac{\delta\mathcal{F}}{\delta\phi} - \partial_k \left(\frac{\delta\mathcal{F}}{\partial_k \delta\phi} \right)\right) \label{eqn:chempot}
\end{align}
and the normalised molecular field
\begin{equation}
    H_{ij}=- \Gamma\frac{\delta\mathcal{F}}{\delta Q_{ij}},
\end{equation}
respectively. Here, $\Gamma_\phi$ is a mobility coefficient and $\Gamma$ indicates the rotational diffusion coefficient, both of which set the rate of relaxation towards the minimum of the free energy $\mathcal{F}\left[\tens{Q},\phi\right]=\int d^2\vec{r} \left( f_Q + f_{\nabla Q} + f_{\phi} + f_{\nabla\phi} \right)$. It can be seen that this free energy includes both bulk and gradient contributions in terms of the tensor order parameter $\tens{Q}$ and the binary order parameter $\phi$, which read
\begin{eqnarray}\label{eq:Blow free energy}
f_{Q}=&\frac{1}{2}\mathcal{C}\left(\phi S_n-2Q_{ij}Q_{ij}\right)^2,\\ 
f_{\nabla Q}=&\frac{1}{2}L\partial_k Q_{ij}\partial_k Q_{ij},\\ 
f_\phi=&\frac{1}{2}\mathcal{A}\phi^2\left(1-\phi\right)^2,\\ 
f_{\nabla\phi}=&\frac{1}{2}K\partial_k\phi\partial_k\phi,\label{eq:Blow free energyEnd}
\end{eqnarray}
where $S_n$ denotes the equilibrium value of the orientational order parameter $S$, and $\mathcal{C}, L, \mathcal{A}, K$ are model parameters. For a more detailed discussion of the model, as well as the used parameter values, see the SI.

The evolution of the binary and orientational order parameters is subsequently completed by means of a coupling to the evolution of the velocity field $\vec{v}$ that is described by the generalised, incompressible Navier-Stokes equations
\begin{equation}\label{eq:NS full}
    \begin{split}
        \partial_iv_i&=0,\\
        \rho\left(\partial_t+v_k\partial_k\right)v_i&=\partial_j\Pi_{ij},
    \end{split}
\end{equation}
 where $\Pi_{ij}$ describes the full stress tensor that includes pressure and is given in the SI.  The dominant terms in the stress tensor are the viscous and capillary stress, which depend on the viscosity of the fluid $\eta$ and the surface tension  $\sigma$  of the binary order parameter $\phi$ \cite{doostmohammadi2016stabilization}. We point out that, in our simulations, the surface tension is no independent model parameter, but rather it is accessed in terms of existing model parameters through the expression
 \begin{equation}
    \sigma=\frac{1}{6}\sqrt{\left(K+\frac{1}{2}L\right)\left(\mathcal{A}_{\text{binary}}+\mathcal{C}_{\text{LQ}}\right)},
\end{equation}
which can be derived by considering the free-energy cost of the interface (see the SI for details). Similar surface tensions have been calculated for solely $\phi$-dependent free energies~\cite{Cahn1958}, but here we have expanded the description to account for the $Q$ dependence of the free energy explicitly.
 For simplicity, we assume surface properties are equal between the outer and inner surfaces. 

Importantly, we also take into account the effect of active stress generated within the cortex. While the activity of the cells is clearly manifested in the brain organoid experiments in the form of active contraction of cells within the cortex~\cite{gundersen2013,karzbrun2018human}, rather surprisingly, to our current knowledge, no prior work has explored the impact of this apparent activity in the dynamics and morphology of the organoids. To account for this contractile activity, we introduce an active stress in the form of coarse-grained stresslets that represent the contractile force dipoles that are generated by the actomyosin machinery of the cell cytoskeleton~\cite{ramaswamy2010mechanics,simha2002hydrodynamic,goriely2017} (see Figure \ref{fig:organoidmodel}b). For simplicity, and to show the generic effect of including active stress we assume a uniform spread of dipoles throughout the organoid, resulting in uniform active stress throughout the active ring. 
Coarse-graining over the dipolar force fields leads to an additional, \textit{active} contribution to the stress tensor~\cite{simha2002hydrodynamic,goriely2017}
\begin{equation}
    \Pi_{ij}^{active}=\alpha Q_{ij},
\end{equation}
with $\alpha$ a proportionality constant  scaling with activity. The sign of $\alpha$ determines the nature of the active stress, with $\alpha > 0$ for contractile and $\alpha < 0$ for extensile active stress. Due to the contractility of the stem cells, we use $\alpha > 0$ in the case of the model brain organoid considered here. However, the activity-induced instability is a generic mechanism that is also expected for extensile activities corresponding, for example, to stress generation due to cell division events~\cite{volfson2008biomechanical,doostmohammadi2015celebrating}.

 The simulations start with zero velocity and the director field oriented along the radial direction (Figure \ref{fig:Initial_instability}a), corresponding to the radially elongated cells between the lumen and the outer interface of the organoid observed in the experiments ~\cite{karzbrun2018human}. Unless otherwise specified, the active region is initialised with the inner interface at radius $a=45$ and outer interface at radius $b=170$.

\section{Simulation results}

\begin{figure*} 
    \centering
    \includegraphics[width=0.95\textwidth]{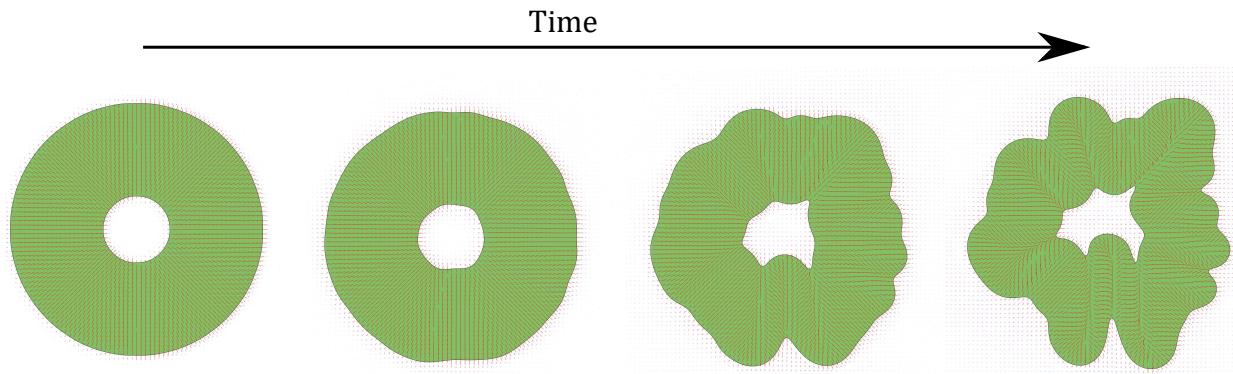}
    \caption{
\textbf{Temporal evolution of the surface folding.}
The modelled cortex region starts as a circle (green denotes the organoid region $\phi=1$), but due to the onset of an active nematic instability, the surfaces fold over time for high activity (depicted here is $\alpha=0.0008$). Displayed timesteps are (0,100000 150000,175000) LB times from left to right. Red lines illustrate modelled cell nuclei direction $q$.
}
\label{fig:Initial_instability}
\end{figure*}

\begin{figure*} 
    \centering
    \includegraphics[width=0.95\textwidth]{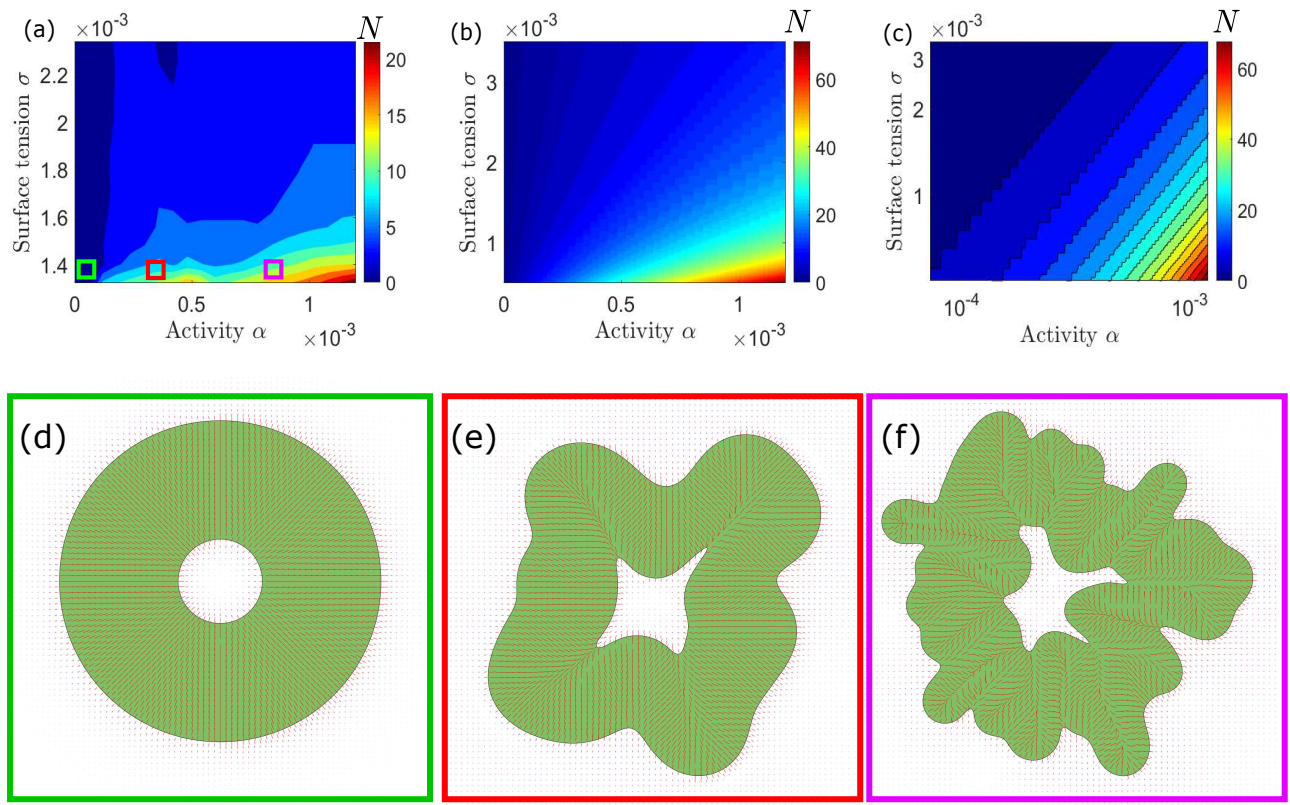}
    \caption{
\textbf{ Folding transition through an increase of contractility. }
The folding number $N$ for different activities and surface tensions found from simulations (a) and the linear stability analysis  in a linear (b) and log (c) scale. 
Simulation snapshots of the organoid for  surface tension 
$\sigma=0.0014$ and activities $\alpha=0.000025$ (d), $\alpha=0.0003$ (e) and $\alpha=0.0008$ (f). 
(e-f) are taken just before formation of first topological defects. 
}
\label{fig:colourmap}
\end{figure*}


We investigate the onset of the interface instability, which resembles gyrification (Figure~\ref{fig:Initial_instability}), by determining the number of folds that form on the initially circular interface of the active ring.
We measure the number of folds (also referred to as the folding or wrinkling number $N$~\cite{karzbrun2018human}) by first measuring the radial distance of the points on the outer interface from the center of the ring.
The number of folds is then determined from the number of maxima in the radial distance signal as a function of azimuthal coordinates at the onset of the instability.
While small perturbations dominate this initial amplitude signal, we notice that a well-defined number of peaks is established relatively quickly. This number remains constant until secondary, non-linear effects begin to dominate the bulk active system, leading to the nucleation of topological defects and the emergence of active turbulence~\cite{thampi2014instabilities,doostmohammadi2018active}. To establish the role of activity in the interface instability and determine the number of folds we focus only on the time span before the bulk instability and the creation of topological defects.

We begin by characterising the number of folds for varying activity and surface tension. The results are represented in the stability diagram (Figure ~\ref{fig:colourmap}a), which clearly demonstrates the competition between the destabilising effect of active stress and the stabilising impact of the surface tension. Increasing activity results in a larger number of folds on the interface, while larger surface tension suppresses the instability and leads to a smaller number of folds. The latter phenomenology is in line with predictions of a purely elastic model of a brain organoid~\cite{riccobelli2020}.
Interestingly, no folding is observed for sufficiently small activities, indicating that there exists a threshold for the active stress exerted by the cells in order to create folds on the interface. 
This distinguishes the interface instability in the ring geometry from the well-known thresholdless hydrodynamic instability of unconfined active nematics~\cite{simha2002hydrodynamic}. More importantly, since we expect the activity to increase with cell nuclear density in the cortex, we conjecture that this observation explains the experimental results of Ref. \cite{karzbrun2018human}, in which no  folding was observed for cell nuclear densities below a critical threshold. Together, these results show that active stress alone can result in hydrodynamic instability of a model organoid, suggesting that activity-induced instabilities can provide a previously overlooked, generic and complementary mechanism to the differential growth mechanism to govern the emergence of folds on brain organoids.



\begin{figure} 
    \centering
    \includegraphics[width=0.45\textwidth]{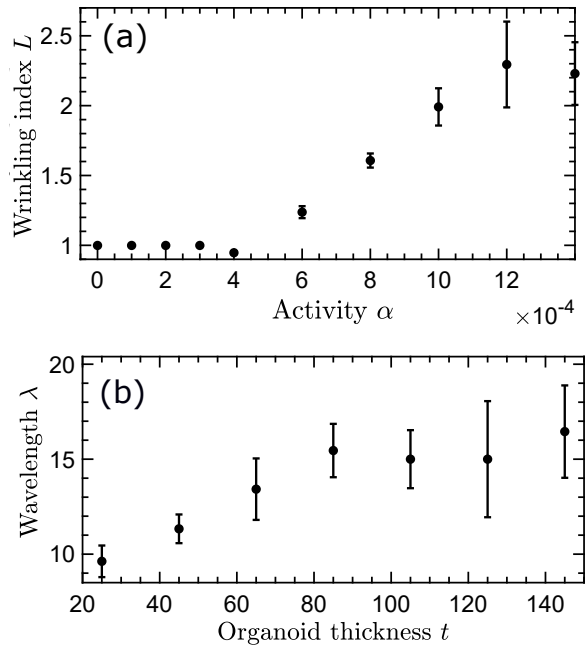}
    \caption{
\textbf{Quantification of the role of activity and organoid thickness.}
(a) The wrinkling index as a function of activity. For low activity, the wrinkling index is 1 (circular) while it increases linearly after a contractile stress threshold. Measurements are performed at a fixed simulation time $t=300000$. (b) Wavelength (defined as the inverse of the  folding number times the outer radius) for different organoid thicknesses, where the inner radius $a$ is varied while retaining the same outer radius $b$.
}
\label{fig:Sim2}
\end{figure} 
To draw more parallels between the experimental results on brain organoids and the active nematic ring model presented here, we reproduce two measurements of the experimental paper  \cite{karzbrun2018human}. 
First, we measure the wrinkling index $L$ of the active nematic ring for different activities at a fixed time (see Figure \ref{fig:Sim2}a).
The wrinkling index is a measurement of the curvature of the interface, defined as the ratio of the contour length normalised to the length of the maximally-protruding outer convex contour. With this definition, $L=1$ denotes a perfect circle without any folding. 
We find that for low activities, the wrinkling index remains 1, indicating the active ring remains a circle and no wrinkles are detected. 
If we increase the activity further, the wrinkling index starts to increase linearly, similar to the linear increase with nuclear cell density found in experiments. 

We also measure the effect of changing the active nematic ring thickness $t$. 
The experiments of Karzbrun et al.~\cite{karzbrun2018human} demonstrated that treating the organoid with blebbistatin resulted in a change in organoid thickness and that the wavelength between different convective wrinkles depends on the thickness. 
We mimic this set-up by varying the inner radius of the isotropic cavity (lumen), while keeping the outer surface radius constant, and define the wavelength $\lambda$ as the contour length over the folding number $N$.
In line with the experimental results, the wavelength $\lambda$ increases with the thickness $t$ up until a certain thickness threshold. After this, the wavelength becomes independent of the thickness (see Figure \ref{fig:Sim2}b). 
Together, these numerical results show that modelling brain organoids as a rings of active nematics can replicate several experimental observations. Next, in order to gain more insight into the nature of the activity-induced instability of the interface, we present a linear stability analysis of the governing equations of the model system.

\section{Linear stability analysis}

The simulation results point to a possible role of activity in driving the interface instability. To provide a better understanding of the possible activity-induced instability, we perform a linear stability analysis on the surfaces of the model organoid at inner radius $a$ and outer radius $b$. For simplicity we consider sharp interfaces, and retain identical surface tensions $\sigma_a=\sigma_b=\sigma$. Furthermore, in order to allow for further analytical treatment we perform the stability analysis in the limit of overdamped friction. 
The active forces that put the organoid surfaces under contractile stress originate at the perturbed surfaces, as---in line with experimental observations~\cite{karzbrun2018human}---we assume the orientation field of the cells retains perpendicular alignment to the interfaces. This corresponds to the limit of strong active anchoring \cite{blow2014biphasic}.
The perturbations of the interfaces thus \textit{directly} induce nematic distortions, which in turn give rise to active forces.
Through this mechanism, we probe the model organoid surfaces for instabilities that resemble gyrification by applying infinitesimal sinusoidal perturbations of the inner and outer interface of the ring of the form:
\begin{eqnarray}
a(t)&=a_0+\delta_a(t)e^{in\theta},\\ b(t)&=b_0+\delta_b(t)e^{in\theta},
\end{eqnarray}
where $\delta(t)$ denotes the infinitesimal perturbation amplitude, $n$ is an integer wavenumber and $\theta$ denotes the azimuthal coordinate.
The wavenumber with the fastest-growing instability is the folding number $N$ that corresponds to the number of folds observed numerically from random perturbations (Figure \ref{fig:colourmap}b-c).


We take the expansion around the quiescent state $\overline{v}_r=0$, $\overline{v}_\theta=0$ and $\overline{p}(r)=\int \diff r\,\alpha S/r$.
The perturbations to the surface then result in perturbative corrections to the velocity of the form $v_r(r,t)=\overline{v}_r+R(r,t)e^{in\theta}$ and $v_\theta(r,t)=\overline{v}_\theta+\Theta(r,t) e^{in\theta}$, as well as a perturbative correction to the pressure $p(r,t)=\overline{p}(r)+P(r,t)e^{in\theta}$. 
 The perturbations evolve according to the Navier-Stokes equations, Eqn. \eqref{eq:NS full}, where we include only the isotropic and active contributions to the stress tensor. Discarding terms of higher order than linear in perturbation, we find
\begin{equation}\label{NS}
\begin{cases}
\begin{aligned}
&\partial_tR=-\frac{1}{\rho}P'+\frac{\alpha S}{\rho}\frac{n^2-1}{b-a}\left\{\frac{b-r}{a^2}\delta_a+\frac{r-a}{b^2}\delta_b\right\}\\
& \; \; \; \; \; \; \; \; \; \: -\chi R\\
&\partial_t\Theta=-\frac{in}{\rho}\frac{P}{r}-\chi \Theta\\
&R'+\frac{R}{r}+in\frac{\Theta}{r}=0,
\end{aligned}
\end{cases}
\end{equation}
where $\rho$ denotes the density of the active nematic, we introduced the overdamped friction coefficient $\chi$ to model viscous effects using an argument similar to Stokes' law \cite{batchelor2000introduction} (see SI) and primes denote derivatives with respect to the radial coordinate $r$. The last term on the first line represents the active forces induced by the perturbed surfaces, which we assume decay linearly away from the surface.

Next, we search for exponential solutions of the form $\propto e^{\omega t}$, where we tacitly assume both surfaces exhibit the same growth rate $\omega$ \cite{dumbleton1970capillary}. This presupposes a hydrodynamic coupling between the organoid surfaces, in accordance with our numerical simulations, and rationalises the dependence of the wrinkling wavelength $\lambda$ on the organoid thickness (see Figure \ref{fig:Sim2}b). In the SI we make this coupling explicit.

Upon inserting this \textit{ansatz}, Eqn. \eqref{NS} reduces to a set of ODEs for the perturbative velocities and pressure, which we solve sequentially. Subsequently demanding continuity and stress balance at the surfaces of the model organoid (see SI) yields a dispersion relation for the growth rate $\omega$:
\begin{equation}\label{eq:overdamped dispersion}
    \omega^4+\frac{2\chi}{\rho}\omega^3+A\omega^2+\chi B\omega+C=0,
\end{equation}
where the coefficients $A$, $B$, $C$ embed dependency on the wave number and model parameters (see the SI for explicit form of these terms).

To determine the stability of a perturbation with wavenumber $n$, we inspect the real part of the corresponding growth rate $\omega$ and associate the wrinkling number $N$ with the fastest-growing mode.
Interestingly, numerical checks of our analytical results show that all unstable surface perturbations are of the same sign, indicating an undulation mode in accordance with experiments \cite{karzbrun2018human}. Furthermore, we recover a clear  folding transition, as---for a given surface tension $\sigma$---non-zero wrinkling numbers are only recovered past a critical activity-to-surface tension ratio (see Figure \ref{fig:colourmap}b-c and the SI). This is in agreement with the numerical simulations found in Figure \ref{fig:colourmap}a, as well as with the experiments of Ref. \cite{karzbrun2018human}, where no wrinkling instability is observed for low cell nuclear densities.

\section{Conclusion}

We present an active matter model system for describing the experimentally-observed folds on brain organoids. By treating the organoid as a contractile, active, biphasic annulus, we demonstrate a generic instability mechanism for the interface deformation due solely to active stress generation by the cytoskeleton of the cells within the active cortex that surrounds a passive lumen without introducing growth, swelling, or any other planar surface forces. Combining numerical simulations with linear stability analyses, we show that the activity-induced instability mechanism occurs over a well-defined activity threshold that depends on the mechanical properties of the system, such as surface tension and the thickness of the model organoid. Since the active stress generation is expected to increase with increasing cell nuclear density within the cortex, these results are in line with the experimental measurements that establish the emergence of folds on the organoid surface above a critical nuclear density~\cite{karzbrun2018human}.

Furthermore, the increase in  folding number $N$ upon increasing activity $\alpha$ (see Figure \ref{fig:colourmap}) lends credence to the proposed model in  light of molecular perturbation experiments on blebbistatin-treated brain organoids that show a marked decrease in the number of folds the organoids exhibit upon treatment with the contractility-inhibiting drug \cite{karzbrun2018human}. 

The activity-induced mechanism proposed here complements previously suggested mechanisms of differential swelling between the inner and outer cortex~\cite{balbi2018mechanics}, adhesion-induced surface tension~\cite{riccobelli2020}, and the fluid-elastic fibrous model~\cite{engstrom2018} for the generation of organoid surface deformations. 

It is important to note that the model presented here provides a generic framework based on accounting for (i) microstructural complexity of the organoid cortex based on the orientational order of the cells, (ii) the biphasic nature of the organoid based on an active, viscoelastic cortex surrounding a fluid-like lumen region, and (iii) active stress generation by the cells. As such, the model allows for isolating the effects of active stress generation from potential impacts of growth-induced mechanisms on the formation of surface instabilities. A more complete picture of the relative importance of different mechanisms leading to brain organoid deformation should allow for activity-induced, growth-induced, and differential adhesion-induced mechanisms to be accounted for in one unifying framework, which should be the focus of future studies.

\section{Acknowledgements}
We thank Alain Goriely, Paul van der Schoot, and Julia M. Yeomans for illuminating discussions. This project has received funding from the European Union's Horizon 2020 research and innovation programme under the Lubiss the Marie Sklodowska-Curie Grant Agreement No. 722497 and the EPSRC grant EP/T031247/1 for K.T. G.L.A.K. acknowledges funding from the Dutch Research Council (NWO) in the framework of the ENW PPP Fund for the top sectors and from the Ministry of Economic Affairs in the framework of the `PPS-Toeslagregeling'. A. D. acknowledges support from the Novo Nordisk Foundation (grant No. NNF18SA0035142), Villum Fonden (Grant no. 29476), Danish Council for Independent Research, Natural Sciences (DFF-117155-1001), and funding from the European Union’s Horizon 2020 research and innovation program under the Marie Sklodowska-Curie grant agreement No. 847523 (INTERACTIONS).

\bibliographystyle{ieeetr}
\bibliography{Bibliography}

\end{document}